\documentclass{IAU-TrA}
\usepackage{graphicx}

\title[SOLAR ACTIVITY]     
{}

\author[DIVISION~II / COMMISSION 10]   
{}

\pubyear{2009}
\volume{Volume XXVIIA}
\pagerange{001--008}
\jname{Transactions IAU, Volume XXVIIA}
\editors{Karel A. van der Hucht, ed.}
\begin{document}

\maketitle

{\bf

\large
\begin{tabbing}
\hspace*{65mm}       \=                                              \kill
COMMISSION~10         \> SOLAR ACTIVITY                            \\[0.5ex]
                     \> {\small\it ACTIVIT\a'{E} SOLAIRE}             \\
\end{tabbing}

\normalsize

\begin{tabbing}
\hspace*{65mm}       \=                                              \kill
PRESIDENT            \> James A. Klimchuk                                \\
VICE-PRESIDENT       \> Lidia van Driel-Gesztelyi                        \\
SECRETARY            \> Carolus J. Schrijver                   \\
PAST PRESIDENT       \> Donald B. Melrose                         \\
ORGANIZING COMMITTEE \> Lyndsay Fletcher, Nat Gopalswamy,              \\
                     \> Richard A. Harrison, Cristina H. Mandrini,          \\
                     \> Hardi Peter, Saku Tsuneta,           \\
                     \> Bojan Vr\v{s}nak, Jing-Xiu Wang                                      \\
\end{tabbing}

\bigskip

\noindent
TRIENNIAL REPORT 2006-2009
}

\firstsection 

\section{Introduction}
Commission 10 deals with solar activity in all of its forms, ranging
from the smallest nanoflares to the largest coronal mass ejections.
This report reviews scientific progress over the roughly two-year
period ending in the middle of 2008.  This has been an exciting time
in solar physics, highlighted by the launches of the Hinode and
STEREO missions late in 2006. The report is reasonably
comprehensive, though it is far from exhaustive. Limited space
prevents the inclusion of many significant results. The report is
divided into the following sections: Photosphere and Chromosphere;
Transition Region; Corona and Coronal Heating; Coronal Jets; Flares;
Coronal Mass Ejection Initiation; Global Coronal Waves and Shocks;
Coronal Dimming; The Link Between Low Coronal CME Signatures and
Magnetic Clouds; Coronal Mass Ejections in the Heliosphere; and
Coronal Mass Ejections and Space Weather.  Primary authorship is
indicated at the beginning of each section.

\section{Photosphere and Chromosphere (C. J. Schrijver)}

\subsection{Quiet-Sun Field Within the Photosphere}

The Solar Optical Telescope (SOT) on board the Hinode spacecraft
provides an unprecedented combination of spatial resolution and
continuity of observations. The SpectroPolarimeter focal-plane
instrumentation exploits that to measure the polarization signals
from the photospheric plasma. Lites et al. (2008) and Ishikawa et
al. (2008) show direct evidence that much of the magnetic field in
the quiet-Sun photosphere is essentially horizontal to the solar
surface.  This observation is the direct confirmation of the
existence of the weak field for which less direct evidence had been
found by Harvey et al. (2007), and contributes to Hanle
de-polarization effects discussed by, e.g., Trujillo Bueno et al.
(2004).

The nearly vertical component is found primarily in the downflow
network of the granular convection, corresponding to the well-known
network field. The horizontal field is mostly found in the interior
of the convective cells. Despite this significant preference for a
separation by upflow and downflow domains, flux has been observed to
also emerge already largely vertical even within the interior of the
granular convective cells (Orozco-Suarez et al. 2008). And, perhaps
not surprisingly, the conceptually expected evolutionary pattern of
emerging flux is also seen: Centeno et al. (2007) report on
observations in which field is seen to first surface nearly
horizontally and subsequently---as it is advected to the downflow
lanes---rights itself to be nearly vertical.

Lites et al. (2008) measure the mean flux density of the horizontal field to be about five times higher than that associated with the nearly vertical field component.
Interestingly, radiative MHD simulations of near-surface stratified convection by Schuessler and Voegler (2008) show a very similar orientation-dependent ratio for the field. Steiner et al.
(2008) reach a similar conclusion based on their numerical
experiments: they argue that the granular upflows allow field to be stretched horizontally, being advected from over the cell centers only slowly in the stagnating, overshooting, upper-photospheric flows. Both studies support the conclusion that near-surface turbulent dynamo action significantly contributes to the internetwork photospheric field. A study by Abbett (2007) elucidates how such a turbulent-dynamo field would connect sub-photospheric and coronal layers through a complex and dynamic chromospheric layer in between; work by Isobe et al. (2008) explores numerically the frequent reconnective interactions expected with the overlying chromospheric canopy field, suggesting that this and the associated wave generation may have significant consequences for atmospheric heating and driving of the solar wind.

\subsection{The Solar Dynamo(s): Global and Local Aspects}

Ephemeral bipolar regions are at the small end of the active region spectrum. Their properties over the solar cycle are an extension of those of their larger counterparts: they follow the general butterfly pattern, and have the proper preferential orientation of their dipole axes relative to the equator, but with a spread about the mean that increases towards the smaller bipoles. In this, they are a natural extension of the active region population. Where they were known to differ from the large regions is in the fact that they are the first to appear and last to fade for a given sunspot cycle.

Now, work by Hagenaar et al. (2008) uncovers another distinct
property of ephemeral regions: the emergence frequency decreases
with increasing local flux imbalance (consistent with findings by
Abramenko et al. (2006) and Zhang et al. (2006) who differentiated
only coronal hole regions from other quiet-Sun regions). Hagenaar et
al. (2008) find that the rate of flux emergence is lower within
strongly unipolar network regions by at least a factor of 3 relative
to flux-balanced quiet Sun. One consequence of this is that because
coronal holes overlie strong network regions, there are fewer
ephemeral regions, and therefore fewer EUV or X-ray bright points
within coronal holes.

The ephemeral-region population thus takes an interesting position
in the study of solar magnetic activity: with the smallest-scale
internetwork field perhaps largely generated by a local turbulent
dynamo (Schuessler \& Voegler 2008), and with the active regions
associated with a global dynamo action, the ephemeral region
population has signatures of both. Voegler \& Schuessler (2007) show
that local dynamo action can lead to a mixed-polarity field similar
to the flux balanced very-quiet network field. It remains to be seen
what such experiments predict in case there is a net flux imbalance,
i.e., a background `guide field': are fewer ephemeral regions
generated, or does reconnection with the background guide field
cause fewer of them to survive the rise to the surface (see
discussion by Hagenaar et al. 2008).

\subsection{Emerging Flux: Observations and Numerical Experiments}

Observations made with the Hinode SOT show unambiguously that
magnetic flux bundles that form active regions do not emerge as
simply curved arches, but rather as fragmented collections of
undulating flux bundles. Each bundle likely crosses the photosphere
one or more times between the extremes of the emerging region (e.g.,
Lites 2008). This is likely the result of the coupling to the
near-surface convective motions, and the difficulty of relatively
heavy sub-photospheric material to drain from the dipped field
segments. Reconnection between neighboring supra-photospheric flux
bundles could pinch off the sub-photospheric mass pockets, thus
allowing the field to rise into the corona. Radiative MHD
simulations of emerging flux by Cheung et al. (2008) support this
interpretation: they show the `serpentine' nature of the emerging
flux, with characteristics that resemble the observed patterns of
emerging flux, flux cancellation with associated downflows,
convective collapse into strong-field flux concentrations, and
photospheric bright points. Note that an example of field dipping
into sub-photospheric layers is also discussed by Abbett (2007).

\subsection{Upper-Chromospheric Dynamics: Spicules and Waves}

Spectacularly sharp Ca II H narrow-band filter observations made
both with Hinode's SOT and the Swedish Vacuum Solar Telescope reveal
ubiquitous jet-like features (called spicules or fibrils) above the
solar limb. The relatively long-lived, broad population among these
(discussed by De Pontieu et al. 2007a) appear to be caused by
acoustic shock waves propagating upward from the photosphere. These
shock waves cause the chromospheric material to undulate with almost
perfectly parabolic height-time profiles, and saw-tooth velocity
patterns. These shock-induced fibrils occur both in plages and in
quiet Sun.

A more enigmatic phenomenon is formed by the much finer and more
transient population of hair-like high extension of the
chromospheric plasma discussed by De Pontieu et al. (2007b). Their
origin remains subject to debate, but their transverse displacements
point to the ubiquitous existence of Alfven-like waves propagating
into the corona.  This is the most direct observational evidence for
the existence of such waves reported to date.  The estimated power
suffices to heat the quiet-Sun corona and power the solar wind:
these waves have amplitudes of 10-25 km s$^{-1}$ for periods of
100-500 seconds. Alfven-like waves with similar periods have also
been observed for the first time in coronal loops, using the CoMP
instrument (Tomczyk et al. 2007).

\section{Transition Region (H. Peter)}

The transition region from the chromosphere to the corona,
originally thought of as a simple thin onion shell-like layer, is a
spatially and temporally highly complex part of the solar
atmosphere. So far we are missing a unifying picture combining the
numerous phenomena observed in  emission lines formed from a couple
of 10,000 K to several 100,000 K.  Some of the aspects are
re-interpreted by Judge (2008), who attempts to  explain the
transition region as being due to cross-field conduction of  neutral
atoms. The emission measure increasing towards low temperatures  and
the persistent redshifts are two of the major observational facts to
be explained. A collection of one-dimensional transient models
(Spadaro et al. 2006) and a three-dimensional MHD model (Peter et
al. 2006) gave quantitative explanations for this. In agreement with
the latter model, Doschek (2006) could show that the bulk of the
(low) transition region emission originates from small cool loops.
Using rocket imaging data, Patsourakos et al. (2007) give further
direct observational evidence for the existence of such small
loop-like structures dominating the transition region emission.

The magnetic field has to connect the transition region structures
down to the chromosphere, and in case they are part of hot coronal
elements also to the corona (Peter 2007). However, correlations
between the transition region and the photosphere cannot be
identified in a unique way (S\'{a}nchez Almeida et al. 2007). Larger
features in the photosphere, such as moving magnetic features, might
well leave an imprint in the outer atmosphere (Lin et al. 2006). In
general, the connection from the chromosphere to the transition
region is quite subtle and hard to identify in observations
(Hansteen et al. 2007). A new way to investigate the relation
between transient events in the transition region and the
chromosphere was presented by Innes (2008). She studied the
chromospheric emission of molecular hydrogen near 111.9 nm during
microflaring events and proposes that the (coronal) energy
deposition in the microflare also heats the chromosphere and thus
affects the opacity for molecular hydrogen lines.

The energy balance, one part being the heating process, is largely
determining the pressure of the transition region and thus
implicitly also the mass loading. Combining various models and
observations, Aschwanden et al. (2007) argue that the bulk part of
the heating is located deep down, basically reflecting an
exponential decay of the heating rate with height, on average. As
speculated earlier, Tian et al. (2008a) could now show that the
persistent blueshifts in upper transition region lines are not due
to the solar wind outflow, but due to mass loading of loops. Recent
Doppler shift observations with the Extreme ultraviolet Imaging
Spectrometer (EIS; Culhane et al. 2007a) on board Hinode indicate
that the redshifts are due to radiative cooling and subsequent bulk
downflows within the loops (Bradshaw 2008).

New investigations of coronal moss, i.e. the (upper) transition
region footpoint areas of large hot loops, show that in moss regions
the temperature is inversely related to the density (Tripathi et al.
2006). Using comparisons with models, Warren et al. (2008b) show
that in order to understand this moss within the framework of a
steady uniform heating model, one needs to assume that the moss
plasma is not fully filling the volume. However, it remains to be
seen if such a static model is applicable at all, because one might
suspect a spatially varying heating rate (see above), and if the
assumption of static moss is justified.

Motivated by direct magnetic field measurements in the corona
indicating the presence of Alfven waves (Tomczyk et al. 2007) and
observations with SOT on Hinode above the limb, McIntosh et al.
(2008) re-interpreted the widths of transition region lines across
the solar disk. They conclude that the present observations are
consistent with a line-of-sight superposition of Alfvenic
disturbances in small-scale structures. How this relates to the new
finding of Doschek et al. (2007) that the (non-thermal) line widths
are largest not in the brightest parts of an active region but in
dimmer regions adjacent to bright loops remains to be seen. Doschek
et al. (2007) find broad lines related to potential outflow
locations, so maybe this problem hints to different acceleration and
heating mechanisms in open and closed field regions. Another new
difference between (globally) open and closed field regions, was
proposed by Tian et al. (2008b), who find evidence that the
expansion of transition region structures is more rapid in the
coronal holes as compared to the quiet Sun. Dolla \& Solomon (2008)
analyzed line widths above the limb in order to determine the
(kinetic) temperatures of minor ions in presumably open field
regions. They find the smallest mass-to-charge-ratio ions to be the
hottest at a given height, but their analysis remains inconclusive
with regard to supporting or disproving the proposed heating by
ion-cyclotron resonances.

While being in orbit nearly 13 years now, the instruments on board
SOHO, SUMER in particular, still give numerous new valuable results
on the transition region. The EIS instrument on board Hinode covers
wavelengths around 17-21 nm and 24-29  nm. This mainly includes
emission lines formed from 1 to several MK, but also a small number
of lines from the transition region, and allows good density
diagnostics (Feldman et al. 2008). Given the spectral range, the
main science topics are grouped around active region phenomena,
while the transition region can also be investigated (Young et al.
2007). Besides these instruments, which will provide the main source
for observations of transition region lines in the coming years,
rocket experiments complement these data.

\section{Corona and Coronal Heating (J. A. Klimchuk)}

    The past two years have seen considerable progress in understanding the
magnetically-closed corona and how it is heated.  This short report
highlights just some of the important contributions.  Much effort
has been devoted to determining the properties of the heating---how
it varies in time and space and whether it depends on physical
parameters such as the strength of the magnetic field and the length
of field lines.  Some studies have concentrated on individual
coronal loops, while others have addressed active regions as a
whole.  These efforts have both clarified some issues and raised new
questions.

    Let us first consider distinct, measurable loops.  A short history is useful.  For
many years after the Skylab soft X-ray observations, it was thought
that loops are static equilibrium structures maintained by steady
heating. Then came the EUV observations from EIT/SOHO and TRACE.
These revealed that warm ($\sim$ 1 MK) loops are much too dense for
static equilibrium and have super-hydrostatic scale heights.
Modeling efforts showed that the excess densities and large scale
heights could be explained by impulsive heating.  Because of their
temperature response, EIT and TRACE are sensitive to the loops when
they are cooling by radiation, well after the heating has ceased.
The problem is that loops are observed to persist for longer than a
cooling time, so a monolithic model is not viable.  This led to the
suggestion that loops are bundles of thin, unresolved strands.  The
observed high densities, large scale heights, and long lifetimes can
all be explained if the strands are heated at different times by a
storm of nanoflares. Since the strands are in different stages of
cooling, a range of temperatures should be present within the loop
bundle at any given time.  In particular, there should be a small
amount of very hot ($> 5$ MK) plasma.  See Klimchuk (2006) for a
discussion of these points and original references.

    Whether loops are isothermal or multi-thermal has been intensely debated over
the past several years.  Double- and triple-filter observations from
TRACE seem to suggest that the most narrow loops are isothermal
(Aschwanden 2008).  However, it has been demonstrated that many
different thermal distributions, including ones that are broad, can
reproduce the observed intensities, even with 3 filters (Schmelz et
al. 2007a; Patsourakos \& Klimchuk 2007; Noglik \& Walsh 2007).
Spectrometer observations provide far superior plasma diagnostics.
The results here are mixed.   Studies made with CDS/SOHO continue to
find evidence for both isothermal loops and highly multi-thermal
loops (Schmelz et al. 2007b; Cirtain et al. 2007), while studies
made with the new EIS instrument on Hinode find that loops tend to
be mildly multi-thermal (Ugarte-Urra et al. 2008; Warren et al.
2008a). Where temporal information is available, there is clear
evidence that the loops are evolving, but the evolution is generally
slower than expected for radiative cooling.  Loop lifetimes are
extremely important and require further investigation.  A loop
bundle will be only mildly multi-thermal if the storm of nanoflares
is short-lived; however, the observed lifetime of the loop will then
be correspondingly short.  If a loop is observed to persist for much
longer than a cooling time (L\'{o}pez Fuentes et al. 2007), then its
thermal distribution is expected to be broad. More work is needed on
whether the lifetimes and thermal distributions of loops are
consistent. Finally, Landi \& Feldman (2008) have found that one
particular active region is dominated by three distinct
temperatures, which would greatly challenge our understanding if
correct.

    Modeling the plasma properties of whole active regions is a relatively new endeavor.
In addition to providing valuable information on coronal heating,
these research models are the forerunners of eventual operational
models for nowcasting and forecasting the solar spectral irradiance.
This is of great practical value, since the irradiance controls the
dynamics, chemistry, and ionization state of the terrestrial upper
atmosphere and thereby affects radio signal propagation, satellite
drag, etc.  Active region models based on static equilibrium are
able to reproduce the observed soft X-ray emission reasonably well,
but they fail, often miserably, at reproducing the EUV emission. The
model corona is too faint in the EUV (there are no warm loops) and
the moss at the transition region footpoints of hot loops is too
bright. Winebarger et al. (2008) have demonstrated that better
agreement can be obtained in the core of an active region by using a
combination of flux tube expansion and filling factors near 10\%.
Filling factors of this magnitude have been measured in moss with
EIS (Warren et al. 2008b).  Small filling factors are consistent
with the idea of unresolved loop strands. Reale et al. (2007) have
developed a new multi-filter technique using XRT/Hinode data that
reveals considerable thermal structure on small but resolvable
scales.

    Active region models based on impulsive heating are in much better
agreement with observations than are static models.  In particular,
the predicted coronal EUV emission is greatly enhanced (Warren \&
Winebarger 2007; Patsourakos \& Klimchuk 2008a). The predicted moss
emission is still too bright, but these models assume constant cross
section flux tubes, and expanding tubes will improve the agreement,
as they do for static models.  It is also likely that the brightness
of the observed moss is diminished by spicules and possibly by other
cool absorbing material.

    Nanoflare models predict that very hot plasma should be present throughout the
corona, albeit in very small quantities (Klimchuk et al. 2008; this
paper presents a highly efficient IDL code for modeling dynamic
loops and is available upon request).  The intensities of very hot
spectral lines are expected to be extremely faint, due to the small
emission measures and possible also to ionization nonequilibrium
effects (Bradshaw \& Cargill 2006; Reale \& Orlando 2008).
Measureable quantities of very hot ($\sim$ 10 MK) plasma have been
detected outside of flares by the CORONAS-F spectroheliometers
(Zhitnik et al. 2006), RHESSI (McTiernan 2008), and XRT (Siarkowski
et al. 2008; Reale et al. 2008).  The derived differential emission
measure distributions, DEM($T$), are consistent with the predictions
of nanoflare models. The DEM($T$) derived from EIS spectra for $T
\leq 5$ MK are also consistent with the predictions (Patsourakos \&
Klimchuk 2008b). Other tests of the nanoflare idea include emission
line Doppler shifts, broadening, and wing enhancements that are
associated with evaporating and condensing plasma (Patsourakos \&
Klimchuk 2006; Hara et al. 2008; Bradshaw 2008).

Information on the distribution of nanoflare energies can be
inferred from the intensity fluctuations of observed loops (Parenti
et al. 2006; Pauluhn \& Solanki 2007; Parenti \& Young 2008; Sarkar
\& Walsh 2008; Sakamoto et al. 2008; Bazarghan et al. 2008). Proper
flares are known to have a power law energy distribution with an
index $< 2$. Extrapolating to smaller energies implies that
nanoflares cannot heat the corona, as first pointed out by Hudson
(1991).  However, it is now believed that the power law index for
small events is $> 2$, though with a large uncertainty (Benz 2004;
Pauluhn \& Solanki 2007; Bazarghan et al. 2008). Furthermore, the
subset of proper flares that are not associated with CMEs also have
a power law index $> 2$ (Yashiro et al. 2006; see Section 6).  Since
the physics of eruptive events and noneruptive events (nanoflares
and confined flares) is likely to be much different, it is not
surprising that they obey different power laws.

    Thermal nonequilibrium, a phenomenon thought to be important for prominence
formation (Karpen \& Antiochos 2008), may also play an important
role in ordinary loops.  Loop equilibria do not exist for steady
heating if the heating is highly concentrated low in the loop legs.
Instead, cool condensations form and fall to the surface in a
cyclical pattern that repeats on a time scale of hours.  Resolvable
condensations are indeed observed in active regions, but only in a
small fraction of loops (Schrijver 2001). As a possible explanation
for other EUV loops, Klimchuk \& Karpen (2008) appeal to the
multi-strand concept. The individual tiny condensations that occur
within each strand will not be detected as long as the strands are
out of phase. It is encouraging that the models predict excess
densities similar to those of observed EUV loops.  Mok et al. (2008)
report thermal nonequilibrium behavior in their active region
simulations. Hot loops form and cool, but without producing
localized condensations.

    We can summarize the state of understanding as follows.  Much of the
magnetically-closed corona is certainly {\it not} in static
equilibrium, but much of it could be.  A significant
portion---perhaps the vast majority---is heated impulsively or is in
thermal nonequilibrium, or some combination thereof.  Most coronal
heating mechanisms that have been proposed involve impulsive energy
release (Klimchuk 2006; Uzdensky 2007; Cassak et al. 2008; Dahlburg
et al. 2008; Rappazzo et al. 2008; Ugai 2008).  It should be noted,
however, that nanoflares that recur sufficiently frequently within
the same flux strand (on a time scale much shorter than the cooling
time) will produce quasi-static conditions.  It is clear that more
observational and theoretical work is required before the coronal
heating problem will be solved.

\section{Coronal Jets (L. van Driel-Gesztelyi)}

Coronal bright points are often observed to have jets---collimated
transient ejections of hot plasma. Hinode  (Kosugi et al. 2007) can
now study the fine detail of jets which tend to occur preferentially
inside coronal holes, which is consistent with reconnection taking
place between the open magnetic field of the coronal hole and the
closed loop field lines. Observations with the XRT instrument (Golub
et al. 2007) revealed that jets from polar coronal holes are more
numerous than previously thought (60 jets day$^{-1}$, Savcheva et
al. 2007; and even 10 jets hour$^{-1}$, Cirtain et al. 2007). The
EIS instrument (Culhane et al. 2007a) allows direct measurement of
the velocity of jets in the corona for the first time. The
footpoints of the loops are seen to be red-shifted which is
consistent with downflowing cooling plasma following reconnection.
The (blue-shifted) jet is the dominant feature in velocity space but
not in intensity (Kamio et al. 2007). Another new feature of jets is
post-jet enhancement of cooler coronal lines observed by EIS. This
can be explained by the hot plasma in the jet not having sufficient
velocity to leave the Sun and then falling back some minutes later
(Culhane et al 2007b).

XRT observations of jets at the poles have shown mean velocities for
jets of 160 km s$^{-1}$ (Savcheva et al. 2007). Multiple velocity
components were found in jets by Cirtain et al. (2007) in XRT polar
coronal hole data: a spatio-temporal average of about 200 km
s$^{-1}$ as well as a much higher velocity  measured at the
beginning of each jet---with speeds reaching 800 km s$^{-1}$.
Cirtain et al. (2007) interpret this early (and sometimes recurrent)
fast flow as being due to plasma ejected at the Alfven speed during
the relaxation phase following magnetic reconnection.  The mass flux
supplied by about 10 jets per hour occurring in the two polar
coronal holes was estimated to produce a net flux of 10$^{12}$
protons m$^{-2}$ s$^{-1}$ which is only a factor of 10 less than the
current estimates of the average solar wind flux. These small jets
are providing a substantial amount of mass that is being carried
into interplanetary space. A 3D numerical simulation has been
carried out to compare with these observations (Moreno-Insertis et
al. 2008) and is found to be consistent with several key
observational aspects of polar jets such as their speeds and
temperatures.

A study of the 3D morphology of jets became possible for the first
time with stereoscopic observations by the EUVI/SECCHI imagers
(Howard et al. 2008) onboard the twin STEREO spacecraft. The most
important geometrical feature of the observed jets was found to be
helical structures showing evidence of untwisting (Patsourakos et
al. 2008). This is in agreement with the 3D model proposed by Pariat
et al. (2008) with magnetic twist  (untwisting) being the jet's
driver.

\section{Flares (L. Fletcher and J. Wang)}

In this brief review we focus on progress in flare energy build-up
and flare prediction, flare photospheric effects, high energy
coronal sources, non-thermal particles, the flare-CME relationship,
and recent advances with Hinode.

Where is the magnetic free energy stored in a flaring active region?
Using the increasingly robust methods for extrapolating magnetic
fields from vector magnetic field measurements, Schrijver et al.
(2008) find evidence for pre-flare filamentary coronal currents
located $< 20$ Mm above the photosphere and Regnier \& Priest (2007)
show that in a newly-emerged active region the free energy is
concentrated within the first 50    Mm (in an older, decaying region
it resides at higher altitudes). Horizontal shear flows close to the
neutral line prior to large flares (Deng et al. 2006) confirm the
concentration of free energy in a small spatial scale, and Schrijver
(2007) finds that if the unsigned flux within 15 Mm of the polarity
inversion line exceeds $2 \! \times \! 10^{21}$ Mx a major flare
will occur within a day. Though Leka \& Barnes (2007) find that the
probability of flaring has only a weak relationship to the state of
the photospheric magnetic field at any time, single or synthesized
magnetic parameters are being used with some success to quantify
flare probability and productivity.  Georgoulis \& Rust (2007)
introduce the effective connected magnetic field of an active
region, finding that this exceeds 1600 and 2100 G for M- and X-class
flares, respectively, at 95\% probability. LaBonte et al. (2007)
surveyed the helicity injection prior to X-class flares producing a
CME, finding occurrence only if the peak helicity flux exceeds $6 \!
\times \! 10^{36}$ Mx s$^{-1}$. Cui et al. (2007) find that flare
probability increases with active region complexity,
nonpotentiality, and length of polarity inversion line. Impulsive
phase HXR sources are concentrated where the magnetic field is
strong, and where the reconnection rate is high (Temmer et al. 2007;
Jing et al. 2008; Liu et al. 2008).

The last three years have seen effort directed towards understanding
the magnetic and seismic effects of flares near the photosphere.  It
is clear that the photospheric magnetic field changes abruptly and
non-reversibly during the flare impulsive phase (see e.g. Sudol \&
Harvey 2005 for a recent survey). Rapid changes in sunspot structure
have also been detected by Chen et al. (2007) in 40\% of X-class
flares, 17\% of M flares, and 10\% C flares, while Wang (2006) finds
variations in magnetic gradient close to the polarity inversion line
consistent with a sudden release of magnetic shear. The obvious
future task is to analyze vector magnetograms to identify changes in
the `twist' component of the field.

Flare-generated seismic waves, discovered by Kosovichev \& Zharkova
(1998) and amply confirmed in Cycle 23, also show the flare's
photospheric impact. Donea et al. (2006), Kosovichev (2006, 2007)
and Zharkova \& Zharkov (2007) show that flare HXR sources and
seismic sources correlate in space and time. Seismic sources are
associated also with white light kernels, responsible for the
majority of the flare radiated energy and strongly correlated with
HXR sources (Fletcher et al. 2007) but the total acoustic energy is
a small fraction of total flare energy (Donea et al. 2006).
Nonetheless, looking at the cyclic variation of the total energy in
the Sun's acoustic spectrum, Karoff \& Kjeldsen (2008) propose
that---analogous with earthquakes---flares may excite long-duration
global oscillations.

The RHESSI mission has discovered several new types of flare coronal
HXR sources, and we highlight here a hard-spectrum HXR source at
least 150 Mm above the photosphere, with a nonthermal electron
fraction of about 10\% (Krucker et al. 2007b). Hard spectrum
gamma-ray (200-800 keV) coronal sources have also been found,
suggesting coronal electron trapping (Krucker et al. 2008). A
soft-hard-soft spectral variation with time is present in some
coronal sources as well as footpoints (Battaglia \& Benz 2006), and
this may be explicable by a combination of coronal trapping and
stochastic electron acceleration (Grigis \& Benz 2006). A curious
observation with the Owens Valley Solar Telescope of terahertz
emission may come from a compact coronal source of electrons at 800
keV, if the electrons are radiating in a volume with a magnetic flux
density of 4.5 kG (Silva et al. 2007).

Discovering the origin and properties of the flare electron
distribution continues to motivate advanced modeling and
observations. Particle-in-cell simulations, used for some years in
magnetospheric physics, have been harnessed to study acceleration in
coronal magnetic islands produced by magnetic reconnection (Drake et
al. 2006), and wave-particle distributions in current sheet and
uniform magnetic field geometries (Karlicky \& Barta 2007; Sakai et
al. 2006). Flare Vlasov simulations are also being developed (e.g.
Miteva et al. 2007; Lee et al. 2008). Detailed RHESSI HXR
spectroscopy has led to a new diagnostic for the flare electron
angular distribution, based on photospheric HXR albedo (Kontar et
al. 2006a). This diagnostic suggests that electron distributions
might not be strongly downward-beamed in the chromosphere (Kontar et
al. 2006b; Kasparova et al. 2007; though see Zharkova \& Gordovskyy
2006 for an alternative explanation). Xu et al. (2008) studied
RHESSI flares having an extended coronal source, finding evidence
for an extended coronal accelerator. Looking at the larger coronal
context, Temmer et al. (2008) show that peaks in the flare electron
acceleration rate and in the CME acceleration rate are simultaneous
within observational constraints ($\sim$ 5 minutes). RHESSI and WIND
observations suggest that the spectral indices of flare and
interplanetary electrons are correlated, but in a way that is
inconsistent with existing models for flare X-ray emission, and the
number of escaping electrons is only about 1/500th of the number of
electrons required to produce the chromospheric HXR flux (Krucker et
al. 2007a).

However, outstanding questions about the total electron number and
supply in solar flares has prompted various authors to suggest
alternatives to the `monolithic' coronal electron beam picture. For
example Dauphin (2007), Gontikakis et al. (2007) examine
acceleration in multiple, distributed coronal region sites, while
Fletcher \& Hudson (2008) investigate the transport of flare energy
to the chromosphere in the form of the Poynting flux of large-scale
Alfvenic pulses, in strong and low-lying coronal magnetic fields.

The relationship between flares and CMEs continues to be an
important topic.  The general consensus regarding the spatial
correspondence between CME position angle and flare location in the
pre-SOHO era was that the flare is located anywhere under the span
of the CME (e.g., Harrison 2006). However, using 496 flare-CME pairs
in the SOHO era, Yashiro et al. (2008) found that the offset between
the flare position and the central position angle (CPA) of the
associated CME has a Gaussian distribution centered on zero, meaning
the flare is typically located radially below the CME leading edge.
This finding suggests a closer flare-CME relationship as implied by
the CSHKP eruption model.  Many flares are not associated with CMEs.
Yashiro et al. (2006) studied two sets of flares one with and the
other without CMEs.  The number of flares as a function of peak
X-ray flux, fluence, and duration in both sets followed a power law.
Interestingly, the power law index was $> 2$ for flares without
CMEs, while $< 2$ for flares with CMEs.  In flares without CMEs, the
released energy seems to go entirely into heating, which suggests
that nanoflares may contribute significantly to coronal heating (see
Section 4).

The launch of Hinode promises significant advances in flare physics
in the next cycle. Thus far there have only been a small number of
well-observed large flares, but observations of small flares start
to show the combined power of RHESSI and Hinode. For example, Hannah
et al. (2008) find a microflare not conforming to the usual
relationship between flare thermal and nonthermal emission, and
Milligan et al. (2008) show evidence for hot downflowing plasma in
the flare corona, not explained in any existing flare model.
Observational evidence for a new kind of reconnection, called
slip-running reconnection, has been found by Aulanier et al. (2007),
and sub-arcsecond structure in the white light flare sources has
been demonstrated by Isobe et al. (2007). We look forward to the
continued operation of these instruments, and the theoretical
advances that they will bring, in the rise of Cycle 24.

\section{Coronal Mass Ejection Initiation (L. van Driel-Gesztelyi)}

In recent years our physical understanding of CMEs has evolved from
cartoons inspired by observations to full-scale numerical 3D MHD
simulations constrained by observed magnetic fields. Notably, there
has been progress made in simulating CME initiation by flux rope
instabilities as inspired by observed filament motions during
eruption which frequently include helical twisting and writhing
(e.g. Rust \& LaBonte 2005; Green et al. 2007).  Several of these
simulations use the analytical model of a solar active region by
Titov \& D\'{e}moulin (1999) as initial condition. The model
contains a current-carrying twisted flux rope that is held in
equilibrium by an overlying magnetic arcade. The two instabilities
considered as eruption drivers are the ideal MHD helical kink and
torus instabilities. The helical kink instability sets in if a
certain threshold of (flux rope) twist ($\sim$ 2.5 turns  for
line-tied flux ropes) is reached (e.g., T\"{o}r\"{o}k \& Kliem
2005). Above this threshold, twist becomes converted into writhe
during the eruption, deforming the flux rope (or filament) into a
helical kink shape. On the other hand, a current-carrying ring (or
flux rope) situated in an external poloidal magnetic field
($B_{ex}$) is unstable against radial expansion when the Lorentz
self-force or hoop force decreases more slowly with increasing ring
radius than the stabilizing Lorentz force due to $B_{ex}$. Known as
the torus instability, its possible role in solar eruptions has been
examined by Kliem \& T\"{o}r\"{o}k (2006) and Isenberg \& Forbes
(2007). In solar eruptions, the torus instability does not require a
pre-eruptive, highly twisted flux rope, but (i) a sufficiently steep
poloidal field decrease with height above the photosphere and (ii)
an (approximately) semi-circular flux rope shape. Both the helical
kink and torus instabilities may be responsible for initiating and
driving prominence/filament eruptions and thus CMEs. The magnetic
field decrease with height above the filament was shown to be
critical whether a confined eruption or a full eruption occurs as
well as for determining the acceleration profile, corresponding to
fast CMEs for rapid (field) decrease, as it is typical of active
regions, and to slow CMEs for gentle decrease, as is typical of the
quiet Sun (T\"{o}r\"{o}k \& Kliem 2005, 2007; Liu 2008). The latter
means that CMEs from complex active regions with steep field
gradients in the corona are more likely to give rise to fast
CMEs---something that is indeed observed.

More complex CME initiation models involve multiple magnetic flux
systems, such as in the magnetic break-out model (e.g. Antiochos et
al. 1999, DeVore \& Antiochos 2008). In this model, magnetic
reconnection removes unstressed magnetic flux that overlies the
highly stressed core field and this way allows the core field to
erupt. The magnetic break-out model involves specific nullpoints and
separatrices. A multi-polar configuration was also included in the
updated catastrophe model (Lin \& van Ballegooijen 2005), the flux
cancellation model (Amari et al. 2007), and the MHD instability
models (T\"{o}r\"{o}k \& Kliem 2007). In an attempt to test CME
initiation models with special attention to the breakout,
Ugarte-Urra et al. (2007) analyzed the magnetic topology of the
source regions of 26 CME events using potential field extrapolations
and TRACE EUV observations. They found only seven events which could
be interpreted in terms of the breakout model, while a larger number
of events (12) could not be interpreted in those terms. The
interpretation of the rest remained uncertain. On the other hand,
the CME event analyzed by Williams et al. (2005) provided a good
example to indicate that also a combination of several mechanisms,
e.g. magnetic break-out and kink instability, can be at work in
initiating CMEs.

\section{Global Coronal Waves and Shocks (B. Vr\v{s}nak)}

The research on globally propagating coronal disturbances
(large-amplitude waves, shocks, and wave-like disturbances)
continued to be very dynamic. Maybe the most prominent
characteristic of the past triennium was an enhanced effort to
combine detailed multi-wavelength observations with the theoretical
background. The empirical research resulted in a number of new
findings, leading to new ideas and interpretations, whereas the
theoretical research provided a better understanding of physical
processes governing the formation and propagation of global coronal
disturbances.

For the first time the EUV signatures of a global coronal wave (`EIT
waves') were measured at high cadence by EUVI/STEREO, related to the
eruption of 2007 May 19 (Long et al. 2008; Veronig et al. 2008).
Long et al. (2008) reported for the first time the wave signatures
at 304 \AA. Furthermore, they confirmed the idea by Warmuth et al.
(2001) that velocities of EIT waves measured by EIT/SOHO are
probably significantly underestimated due to the low cadence of the
EIT instrument. Veronig et al. (2008) revealed reflection of the
wavefront from the coronal hole boundary, indicating that the
observed disturbance represents a freely-propagating MHD wave.

The data from pre-STEREO instruments continued to be exploited
fruitfully. Mancuso \& Avetta (2008) analyzed the UV-spectrum
(UVCS/SOHO) of the 2002 July 23 coronal shock, and concluded that
the plasma-to-magnetic presure ratio $\beta$   could be an important
parameter in determining the effect of ion heating at collisionless
shocks. Employing the extensive data on CMEs, solar energetic
particle (SEP) events, and type II radio bursts during the SOHO era,
Gopalswamy et al. (2008b) demonstrated that essentially all type II
bursts in the decameter-hectometric (DH) wavelength range are
associated with SEP events once the source location is taken into
account. Shen et al. (2007) proposed a method to determine the shock
Mach number by employing the CME kinematics, type II burst dynamic
spectrum, and the extrapolated magnetic field. Analyzing one Moreton
wave that spanned over almost $360^{\rm o}$, Muhr et al. (2008)
revealed two separate radiant points at opposite ends of the
two-ribbon flare, indicating that the wave was driven by the CME
expanding flanks. Veronig et al. (2006) found out that the
Moreton/EIT wave segments where the front orientation is normal to
the coronal hole boundary can intrude into the coronal hole up to
60-100 Mm.

Regarding the nature of global coronal disturbances, some new ideas
appeared. For example, Attrill et al. (2007) attributed EIT waves to
successive reconnections of CME flanks with coronal loops, which
could explain the association of EIT `waves' and shallow coronal
dimmings which are formed behind the bright front. Wills-Davey et
al. (2007) proposed that slow EIT waves are caused by MHD slow-mode
soliton-like waves. Delannee et al. (2008) performed a 3D MHD
simulation to show that EIT `waves' could be a signature of a
current shell formed around the erupting structure. Balasubramaniam
et al. (2007) demonstrated that the visibility of Moreton waves
increases when sweeping over filaments and filament channels, so
they put forward the idea that a significant contribution to the
Moreton-wave H$_{\alpha}$  signature might be coming from coronal
material of enhanced density.

The question of the origin of coronal shocks and large amplitude
waves continues to be one of the central topics in this field. The
published studies showed a variety of results, some biased towards
the CME-driven option, some favoring the flare-ignited scenario, and
some finding arguments for small scale ejecta (e.g., Chen 2006;
Pohjolainen \& Lehtinen 2006; Shanmugaraju et al. 2006a,b;
Subramanian \& Ebenezer 2006; Cho et al., 2007; Liu et al. 2007;
Reiner et al. 2007; White 2007; Grechnev et al. 2008; Muhr et al.
2008; Veronig et al. 2008; Magdalenic et al. 2008; Mancuso \& Avetta
2008; Pohjolainen 2008). To illustrate the current level of
ambiguity in such studies, let us mention that for one well observed
event two sets of authors came to diametrally opposite conclusions:
Vr\v{s}nak et al. (2006) favored a flare driver, whereas Dauphin et
al. (2006) advocate a CME. The status of the `CME/flare controversy'
was reviewed recently by Vr\v{s}nak \& Cliver (2008).

Related to the formation and propagation of large-amplitude waves
and shocks, a number of important theoretical papers were published.
Pagano et al. (2007) investigated the role of magnetic fields and
showed that a CME-driven wave propagates to longer distances in the
absence of magnetic field than in the presence a weak open field.
Ofman (2007) modeled the wave activity following a flare by
launching a velocity pulse into a model active region and
demonstrated that the resulting global oscillations are in good
agreement with observations. Employing the photospheric magnetic
field measurements, Liu et al. (2008) performed a 3D MHD simulation
of a CME, and showed that the shock segment at the nose of the CME
remains quasi-parallel most of the time. In the simulation of
reconnection in a vertical current sheet, Barta et al. (2007)
revealed the formation of large-amplitude waves associated with
changes of the reconnection rate, which might explain
flare-associated type II bursts in the wake of CMEs. Zic et al.
(2008) developed an analytical MHD model describing the formation of
large-amplitude waves by impulsively expanding 3D pistons. The model
provides an estimate of the time/distance at which the shock should
be formed, dependent on the source-surface acceleration, the
terminal velocity, the initial source size, the ambient Alfven
speed, and plasma $\beta$.

Finally, it should be noted that a comprehensive review on coronal
waves and shocks was published by Warmuth (2007). Gopalswamy (2006e)
reviewed the relationship between CMEs and type II bursts, while
Mann \& Vr\v{s}nak (2007) surveyed the relationship between CMEs,
flares, coronal shocks, and particle acceleration.

\section{Coronal Dimming (R. Harrison and L. van Driel-Gesztelyi)}

There is no strict definition of the phenomenon which we call
coronal dimming. Most authors consider coronal dimming to be a
depletion of extreme-UV (EUV) or X-ray emission from a large region
of the corona, which is thought to be closely associated with
coronal mass ejection (CME) activity. Clearly, understanding the
onset phase of a CME is one of the key issues in solar physics
today, so the study of such dimming activity could well be of
critical importance. However, most of the literature deals with
dimming in a rather hand-waving manner, with the emphasis on
phenomenological studies and associations, no strict definitions of
what constitutes a dimming event (e.g., the depth of the depletion
in intensity, the size of the dimming region, etc.) and little in
terms of a physical interpretation of the plasma characteristics of
the dimming region. Having said that, some key studies are emerging
which do tackle such issues head on, and with the advent of the new
STEREO and Hinode spacecraft, along with the on-going SOHO and TRACE
missions, as well as the up-coming SDO mission, we have many tools
to address this area of research effectively.

Coronal dimming is not a newly discovered phenomenon; Rust and
Hildner (1976) reported such an event using Skylab observations.
More recently, from the late 1990s, dimming was reported using X-ray
and EUV, imaging and spectroscopic data, from the SOHO and Yohkoh
spacecraft (e.g. Sterling \& Hudson 1997; Harrison 1997; Gopalswamy
\& Hanaoka 1998; Zarro et al. 1999; Harrison \& Lyons 2000), and
dimming has taken center-stage in the study of mass ejection onset
in recent years (e.g., recent studies include Moore \& Sterling
2007; Zhang et al. 2007; Reinard \& Biesecker 2008). In many ways
coronal dimming has become a well established phenomenon.

The majority of dimming reports involve EUV or X-ray imaging, and we
have excellent tools aboard SOHO, TRACE, STEREO and Hinode to
identify and study the topology and evolution of dimming regions. On
the other hand, there are spectroscopic studies of dimming which are
providing key plasma information, despite having limited fields of
view or cadence. The combination of imaging and spectroscopy is
essential, but it is worth stressing some of the spectroscopic
studies because they stress the physical processes which are
involved in the dimming and, perhaps, the CME onset process.

EUV spectroscopy has been used to confirm that the dimming process
represents a loss of mass---i.e., it is a density depletion---rather
than a change in temperature (Harrison \& Lyons 2000; Harrison et
al. 2003). Indeed, these studies have demonstrated the loss of
between $4.3 \! \times \! 10^{10}$ and $2.7 \! \times \! 10^{14}$
kg, in each case consistent with the mass of an overlying,
associated CME. If we are identifying the plasma which becomes (part
of) the CME, then this is an exciting phenomenon; studies focusing
on the properties of the dimming plasma, before, during and after
the event, will be essential for understanding the CME onset
(Harrison \& Bewsher 2007).

Hudson et al. (1996) showed that the timescale of the dimming
formation observed in Yohkoh/Soft X-ray Telescope (SXT; Tsuneta et
al. 1991) data is much faster than corresponding conductive and
radiative cooling times.  More recently, data obtained by the
Hinode/Extreme ultra-violet Imaging Spectrometer (EIS; Culhane et
al. 2007) have shown detection of Doppler blueshifted plasma
outflows of velocity $\approx 40$ km s$^{-1}$ corresponding to a
coronal dimming (Harra et al. 2007). This result confirms a similar
finding (Harra \& Sterling 2001) obtained with the SOHO/Coronal
Diagnostic Spectrometer (CDS; Harrison et al. 1995). In addition,
SOHO/CDS limb observations have been used to show the formation of a
dimming region through the outward expansion of pre-CME EUV loops
(Harrison \& Bewsher, 2007), which is consistent with such
blueshifts. Imada et al. (2007) find that Hinode/EIS data of a
dimming shows a dependence of the outflow velocity on temperature,
with hotter lines showing a stronger plasma outflow (up to almost
150 km s$^{-1}$. These works collectively support the primary
interpretation of coronal dimmings as being due to plasma
evacuation.

Statistical studies are becoming important in truly establishing the
relationship with CMEs. Reinard \& Biesecker (2008) have recently
studied the properties of 96 dimming events, using EUV imaging,
associated with CME activity.  They confirmed earlier studies which
showed that the dimming events could be long-lasting, ranging from 1
to 19 hours, and compared the size of the dimming regions to the
associated CMEs. They also tracked the number of dimming pixels
through each event and showed that the `recovery' after the dimming
often took the form of a two-part slope (plotted as dimming area vs.
time).

Bewsher et al. (2008) have produced the first statistical and
probability study of the dimming phenomenon using spectroscopy. They
recognized that while we have associated CMEs and dimming, there has
not been a thorough statistical study which can really identify the
degree of that association, i.e., to put that relationship on a firm
footing. Using spectroscopy, they also recognized the importance of
studying this effect for different temperatures. They made use of
over 200 runs of a specific campaign using the SOHO spacecraft with
an automated procedure for identifying dimming.

Key results included the following:  Up to 84\% of the CMEs in the
data period can be back-projected to dimming events---and this
appears to confirm the association that we have been proposing.
However, they also showed, as did other spectral studies, that the
degree of dimming varies between temperatures from event to event.
If different dimming events have different effects at different
temperatures then this is a problem for monitoring such events with
fixed-wavelength imagers.

Assuming that magnetic field lines of the CME are mostly rooted in
the dimmings, several properties derived from the study of dimmings
can be used to obtain information about the associated CME. Firstly,
calculations of the emission measure and estimates of the volume of
dimmings can give a proxy for the amount of plasma making up the CME
mass (Sterling \& Hudson 1997; Harrison \& Lyons 2000; Harrison et
al. 2003; Zhukov \& Auch\`{e}re 2004). Secondly, the spatial extent
of coronal dimmings can give information regarding the angular
extent of the associated CME (Thompson et al. 2000; Harrison et al.
2003; Attrill et al. 2007, van Driel-Gesztelyi et al.  2008).
Thirdly, quantitative measurement of the magnetic flux through
dimmings can be compared to the magnetic flux of modeled magnetic
clouds (MC) at 1 AU (Webb et al. 2000; Mandrini et al. 2005; Attrill
et al. 2006; Qiu et al. 2007), see D\'{e}moulin (2008) for a review.
Fourth, studying the evolution of the dimmings, particularly during
their recovery phase can give information about the evolution of the
CME post-eruption (Attrill et al. 2006; Crooker \& Webb 2006)
providing proof for e.g. magnetic interaction between the expanding
CME and open field lines of a neigboring coronal hole. Finally,
study of the distribution of the dimmings, their order of formation
and measurement of their magnetic flux contribution to the
associated CME enabled Mandrini et al. (2007) to derive an
understanding of the CME interaction with its surroundings in the
low corona for the case of the complex 28 October 2003 event. They,
building on the model proposed by Attrill et al. (2007),
demonstrated that magnetic reconnection between field lines of the
expanding CME with surrounding magnetic structures ranging from
small- to large-scale (magnetic carpet, filament channel, active
region) make some of the field lines of the CME `step out' from the
flaring source region. Magnetic reconnection is driven by the
expansion of the CME core resulting from an over-pressure relative
to the pressure in the CME's surroundings. This implies that the
extent of the lower coronal signatures match the final angular width
of the CME. Through this process, structures over a large-scale
magnetic area become CME constituents (for a review see van
Driel-Gesztelyi et al. 2008). From the wide-spread coronal dimming
some additional mass is supplied to the CME.

Observations show that coronal dimmings recover whilst suprathermal
uni- or bi-directional electron heat fluxes are still observed at 1
AU in the related ICME, indicating magnetic connection to the Sun.
The questions why and how coronal dimmings disappear whilst the
magnetic connectivity is maintained was investigated by Attrill et
al. (2008) through the analysis of three CME-related dimming events.
They demonstrated that dimmings observed in SOHO/EIT data recover
not only by shrinking of their outer boundaries but also by internal
brightenings. They show that the model developed in Fisk \&
Schwadron (2001) of interchange reconnections between `open'
magnetic field and small coronal loops is applicable to observations
of dimming recovery. Attrill et al. (2008) demonstrate that this
process disperses the concentration of `open' magnetic field
(forming the dimming) out into the surrounding quiet Sun, thus
recovering the intensity of the dimmings whilst still maintaining
the magnetic connectivity of the ejecta to the Sun.

Although this brief summary cannot report on all studies, it is
clear that we have made progress very recently in putting the
dimming phenomenon on a firm footing---the association is real---and
we are making in-roads into studies of the plasma activities leading
to the dimming/CME onset process. With the continuation of the SOHO
mission, as well as TRACE, combined with the new STEREO and Hinode
missions and the up-coming SDO mission, this is a topic which will
receive much attention in the next few years.

\section{The Link Between Low-Coronal CME Signatures and Magnetic Clouds (C. Mandrini)}

    A major step to understanding the variability of the space environment is to
link the sources of coronal mass ejections (CMEs) to their
interplanetary counterparts, mainly magnetic clouds (MCs), a subset
of interplanetary CMEs characterized by enhanced magnetic field
strength when compared to ambient values, a coherent and large
rotation of the magnetic field vector, and low proton temperature
(Burlaga 1995). Identifying the solar sources and comparing
qualitatively and quantitatively global characteristics and physical
parameters both in the Sun and the interplanetary medium provide
useful tools to constrain models in both environments.

    Under the assumption that dimmings (see Section 9) at the Sun mark the
position of ejected flux rope footpoints (Webb et al. 2000), the
magnetic flux through these regions can be used as a proxy for the
magnetic flux involved in the ejection and, thus, be compared to the
magnetic flux in the associated interplanetary MC. Another proxy for
the flux involved in an ejection is the reconnected magnetic flux
swept by flare ribbons, as they separate during the evolution of
two-ribbon flares. Using EUV dimmings as proxies and reconstructing
the MC structure from one spacecraft observations, Mandrini et al.
(2005) and Attrill et al. (2006) found that the magnetic flux in
dimming regions was comparable to the azimuthal MC flux, while the
axial MC flux was several times lower. Qui et al. (2007) analyzed
and compared the reconnected magnetic flux to the total MC flux,
finding similar results (see also Yurchyshyn et al. 2006; Longcope
et al. 2007; M\"{o}stl et al. 2008, where MC data from two
spacecraft were used).
    These results led to the conclusion that the ejected flux rope is formed
by successive reconnections in a sheared arcade during the eruption
process, as opposed to the classical view of a previously existing
flux rope being ejected. However, in extreme events that occur in
not isolated magnetic configurations, it was found that the flux in
dimmings did not agree with the MC flux (Mandrini et al. 2007). This
mismatch led these authors to propose a scenario in which dimmings
spread out to large distances from the initial erupting region
through a stepping reconnection process (in a similar process to
that proposed by Attrill et al., 2007, for the interpretation of EIT
waves). An overview of earlier works on quantitative comparisons of
solar and interplanetary global magnetohydrodynamic invariants, such
as magnetic flux and helicity, can be found in D\'{e}moulin (2008).

    Qualitative comparisons are also useful tools to understand the eruption
process. Studying the temporal and spatial evolution of EUV
dimmings, together with soft X-ray coronal observations, in
conjunction with interplanetary {\it in situ} data of suprathermal
electron fluxes, Attrill et al. (2006) and Crooker \& Webb (2006)
derive an eruption scenario in which interchange magnetic
reconnection between the expanding CME loops and the open field
lines of a polar coronal hole led to the opening of one leg of the
erupting flux rope. Harra et al. (2007), combining EUV and
H$_{\alpha}$ solar observations of eruptive events with {\it in
situ} magnetic field and suprathermal electron data, were able to
understand the sequence of events that produced two MCs with
opposite magnetic field orientations from the same magnetic field
configuration.

    The simple comparison of the magnetic field orientation in the erupting
configurations, which can be inferred from magnetograms, the
directions of filaments, coronal arcades or loops, with the axis of
the associated MCs, can give clues about the mechanism at the origin
of solar eruptions. Green et al. (2007) analyzed in detail
associations of filament eruptions and corresponding MCs, and they
found that when the filament and MC axis differed by a large angle,
the direction of rotation was related to the magnetic helicity sign
of the erupting configuration (see also Harra et al. 2007). The
rotation was consistent with the conversion of twist into writhe,
under the ideal MHD constraint of helicity conservation, providing
support for the assumption of a flux rope topology where the kink
instability sets in during the eruption (see the review by Gibson et
al. 2006).

\section{Coronal Mass Ejections in the Heliosphere (R. Harrison)}

In the 1970s the Helios spacecraft operated from solar orbits with
perihelion 0.31 AU. Zodiacal light photometers were used to detect
CMEs in the inner heliosphere (see e.g. Richter et al. 1982; Jackson
\& Leinert 1985). CME images were constructed from three photometers
which scanned the sky using the spacecraft rotation. More recently,
a major advance was made with the launch, in 2003, of the Solar Mass
Ejection Imager (SMEI) aboard the Coriolis spacecraft (Eyles et al.
2003). This instrument maps the entire sky with three cameras each
scanning $60^{\rm o}$ slices of the sky as the spacecraft moves
around the Earth, and thus, it has pioneered full-sky mapping aimed
specifically at the detection of CMEs propagating through the inner
heliosphere (see e.g. recent papers by Kahler \& Webb 2007 and
Jackson et al. 2007).

The combination of wide-angle heliospheric mapping from out of the
Sun-Earth line is now being satisfied by the Heliospheric Imagers
(HI) (Harrison et al. 2008) aboard the NASA STEREO spacecraft. The
development of these instruments has come very much from the SMEI
heritage and, with the unique opportunities from the STEREO
spacecraft locations, these instruments are able to image those CME
events directed towards the Earth. Indeed, for the first time, the
HI instruments provide a view of the passage of CMEs along virtually
the entire Sun-Earth line and such observations represent a major
milestone in investigations of the influence of solar activity on
the Earth and human systems.

Each HI instrument consists of two wide-angle telescopes mounted
within a baffle system enabling imaging of the heliosphere from the
corona out to Earth-like distances and beyond. The low scattered
light levels and sensitivity allow the detection of stars down to
magnitudes of 12-13. This performance is excellent for the detection
of solar ejecta and solar wind structure through the detection of
Thomson scattered photospheric light off free electrons in regions
of density enhancement.

The STEREO spacecraft were launched in October 2006 with full
scientific operation of the HI instruments starting from April 2007.
The spacecraft are in near Earth-like solar orbits, with one ahead
and one behind the Earth in its orbit. They are drifting away at
$22.5^{\rm o}$ per year (Earth-Sun-spacecraft angle). The spacecraft
are labelled STEREO A and STEREO B, for ahead and behind.

The first HI observations of CMEs in the heliosphere, tracked out to
Earth-like distances, were reported by Harrison et al. (2008). The
same instruments are also reaping the benefits of wide-angle imaging
of the heliosphere with observations of comets (Fulle et al. 2007;
Vourlidas et al. 2007), even the imaging of co-rotating interaction
regions (Sheeley et al. 2008a,b; Rouillard et al. 2008a) and impacts
of CMEs at other planets (Rouillard et al. 2008b).

With the HI instruments we now have a real opportunity to begin to
relate the coronal events that we call CMEs with their heliospheric
counterparts, commonly referred to as ICMEs - Interplanetary CMEs.
Most ICME studies have been performed utilizing {\it in situ}
particle and field observations, and it is clear that heliospheric
imaging can provide a thorough test of the interpretation of such
{\it in situ} data on the topology and propagation of CMEs in the
heliosphere. Indeed, the uniqueness of this opportunity is well
illustrated by the fact that there are a number of extremely basic
observational tests which can be made with the new facility to
underline our current understanding of how CMEs travel out through
the Solar System.

Crooker \& Horbury (2006) have recently reviewed the propagation of
ICMEs in the heliosphere, utilizing {\it in situ} data. They note
that cartoon sketches of ICMEs commonly show magnetic field lines
connected to the Sun at both ends. Furthermore, reporting on the
work of Gosling et al. (1987), Crooker et al. (2002), and others,
they note that it is widely accepted that counter-streaming particle
beams in ICMEs are a sign that both ends of the ICME are indeed
connected to the Sun. This is known as a `closed' ICME. On the other
hand, uni-directional beams may signal connection at only one
end---an `open' ICME. Logically, then, the lack of beams would
appear to signal disconnection at both ends. In this case the ICME
has become an isolated plasmoid.

Given this interpretation, {\it in situ} observations of ICMEs
appear to show many events which are apparently connected to the Sun
at both footpoints, and rather fewer events which appear to be
connected at one end. Complete disconnection of an ICME (a plasmoid)
appears to be rare. In addition, the {\it in situ} observations
suggest that CMEs are connected to the Sun over extremely long
distances; Riley et al. (2004) looked for the degree of `openness'
of ICMEs using observations of counter-streaming electrons from
Ulysses data and could detect no trend in the openness of ICMEs with
distance out to Jupiter. If ICME connectivity to the Sun is the same
at 1 AU as it is at 5 AU then it can be argued that an ascending CME
could still be rooted at the Sun for a week, or, indeed, much
longer.

In reality, an ascending flux rope would most likely contain a mix
of open and closed field lines, driven by apparently random
reconnection events (Crooker \& Horbury 2006; Gosling et al. 1995).
Complete disconnection of the structure appears to be unlikely.

With the new STEREO HI data we should be able to test this scenario,
and this has been reported by Harrison et al. (2008). The HI data
appear to confirm the {\it in situ} interpretation showing coherent
structures, apparently still connected to the Sun over long
distances. There is no evidence for events pinching-off. However,
this in turn presents us with an anomaly. McComas (1995) has argued
that the heliospheric magnetic flux does not continually build up,
so flux must be shed through reconnection somehow during the ICME
process. If we are rejecting the plasmoid or disconnected ICME
scenario then we must find another way of limiting the flux build up
over time.

In the absence of evidence for the pinching-off of CMEs, an
interchange reconnection process has been suggested as the mechanism
by which CMEs disconnect from the Sun (Gosling et al. 1995; Crooker
et al. 2002). The basic idea is that the ascending CME can travel a
considerable distance, well beyond the Earth, still connected to the
Sun, and that perhaps days or even weeks after the onset, the legs
of the CME, still rooted in the Sun, will interact with adjacent
open field lines at low altitude in the corona; reconnection results
in the formation of low-lying loops as one CME leg reconnects with
the adjacent fields and an outward ascending kink-shaped structure
ascends into the heliosphere from the site of one of the original
CME footpoints.

This approach has a few attractive points. For example, it seems
logical that the site of the greatest field density, magnetic
complexity and field-line motion would be the most likely site of
any reconnection in the ascending CME. However, assuming that such
interchange reconnection is the `end game' of a CME, and that this
low level reconnection results in the outward propagation of a
kinked field-line configuration,  what might we expect to observe
and, indeed, have we seen such features? Harrison et al. (2008)
indeed point to observations of narrow V-shaped structures
identified in the HI data that could be candidates for such
reconnection events.

It is early days for this work using STEREO but the indications are
that there is plenty to be gained from these studies. As the mission
progresses we anticipate more opportunities where we have the chance
to combine both imaging and {\it in situ} measurements of specific
events, and their impacts, as well as to model CMEs in the
heliosphere in 3D as never before. Thus, this report should be take
as an early statement on the progress and direction of this work
which is opening a new chapter in solar, heliospheric and space
weather physics.

\section{Coronal Mass Ejections and Space Weather (N. Gopalswamy)}

CMEs cause adverse space weather in two ways: (i) when they arrive
at Earth's magnetosphere, they can couple to Earth's magnetic field
and cause major geomagnetic storms (Gosling et al. 1990) and (ii)
they can drive fast mode MHD shocks that accelerate solar energetic
particles (Reames 1999).  Significant progress has been made on both
these aspects over the past few years. In the case of geomagnetic
storms, connecting the magnetic structure and kinematics of ICMEs
observed at 1 AU to the CME source region at the Sun has received
considerable attention. In the case of SEPs, assessing the
contribution from flare reconnection and shock to the observed SEP
intensity has been the focus. The importance of the variability in
the Alfven speed profiles in the outer corona is also under
investigation because of its importance in deciding the shock
formation.

\subsection{Geomagnetic Storms}

High-Speed Solar Wind Streams (HSS) interacting with the slow solar
wind result in corotating interaction regions (CIRs), which also can
produce geomagnetic storms (Vr\v{s}nak et al. 2007a), but they are
generally weaker than the CME-produced storms (Zhang et al. 2007).
Occasionally, the CIR and ICME structures combine to produce major
storms (Dal Lago et al. 2006).  Multiple CMEs are often involved
producing some super-intense storms (Gopalswamy et al. 2007; Zhang
et al. 2007). There are numerous effects produced by the ICMEs in
the magnetosphere and various other layers down to the ground (see
Borovsky et al. 2006; Kataoka \& Pulkkinen 2008).

The key element of ICMEs for the production of geomagnetic storms is
the southward magnetic field component. While the quite heliospheric
field has no out of the ecliptic field component (except for
Alfvenic fluctuations in the solar wind), a CME adds this component
to the interplanetary (IP) magnetic field. If an ICME has a flux
rope structure, one can easily see that the azimuthal component of
the flux-rope field or its axial component forms the out of the
ecliptic component.  In ICMEs with a flux rope structure (i.e.,
magnetic clouds), it is easy to locate the southward component from
the structure of the cloud (Gopalswamy 2006a; Wang et al. 2007;
Gopalswamy et al. 2008a). In non-cloud ICMEs, it is not easy to
infer the location of the southward component. If the ICMEs are
shock-driving, then the magnetosheath between the shock and the
driving ICME (Kaymaz \& Siscoe 2006; Lepping et al. 2008) can
contain southward field and hence cause geomagnetic storms
(Gopalswamy et al. 2008a). The cloud and sheath storms can be
substantially different (Pulkkinen et al. 2007).

Once an IP structure has a southward magnetic field, the efficiency
with which it causes geomagnetic storm depends on the strength of
the magnetic field and the speed with which it hits the
magnetosphere (Gonzalez et al. 2007; Gopalswamy 2008d). Statistical
investigations have shown that the storm intensity (measured e.g.,
by the Dst index) is best correlated with the speed-magnetic field
product in magnetic clouds and their sheaths. Interestingly, an
equally good correlation is obtained when the magnetic cloud/sheath
speed is replaced by the CME speed measured near the Sun (Gopalswamy
et al. 2008a). This suggests that if one can estimate the magnetic
field in CMEs near the Sun, the strength of the ensuing magnetic
storm can be predicted. The ICME speed can be predicted based on the
CME speed by quantifying the interaction between CMEs and the solar
wind (Xie et al. 2006; Nakagawa et al. 2006; Jones et al. 2007;
Vr\v{s}nak \& Zic, 2007). Most of the storm-causing CMEs are halo
CMEs, which are subject to projection effects and hence space speeds
cannot be easily measured (Kim et al. 2007; Gopalswamy \& Xie 2008;
Howard et al. 2007; Vr\v{s}nak et al. 2007b). There have been
several attempts to use the sky-plane speed of CMEs to obtain their
space speed (Xie et al. 2006; Michalek et al. 2008; Zhao 2008) with
varying extents of success. The magnetic field strength and kinetic
energy of CMEs are somehow related to the free energy available in
the source region. Quantifying this free energy has been a difficult
task (Ugarte-Urra et al. 2007; Schrijver et al. 2008).

The solar sources of CMEs need to be close to the disk center for
the CMEs to make a direct impact on Earth and they have to be fast.
In fact the solar sources of magnetic clouds, storm-causing CMEs,
and halo CMEs have been shown to follow the butterfly diagram
suggesting that only sunspot regions have the ability to produce
such energetic CMEs (Gopalswamy 2008d). The average near-Sun speed
of CMEs that cause intense geomagnetic storms is $\sim 1000$ km
s$^{-1}$ (Gopalswamy 2006b; Zhang et al. 2007), similar to the
average speed of halo CMEs (Gopalswamy et al. 2007) because many of
the storm-producing CMEs are halo CMEs. Halo CMEs are more energetic
(Lara et al. 2006; Liu 2007; Gopalswamy et al. 2007, 2008a) and end
up being magnetic clouds at 1 AU. Most halo CMEs (~70\%) are
geoeffective. Non-geoeffective halos are generally slower, originate
far from the disk center, and originate predominantly in the eastern
hemisphere of the Sun.  The geoeffectiveness rate of halo CMEs has
been reported to be anywhere from ~40\% to more than 80\% (Yemolaev
\& Yermolaev 2006), but the difference seems to be due to different
definitions used for halo CMEs (some authors have included all CMEs
with width $> 120^{\rm o}$ as halos) and the sample size (Gopalswamy
et al. 2007).  The geoeffectiveness rate of CMEs may be related to
the fact that more ICMEs are observed as magnetic clouds during
solar minimum than during solar maximum (Riley et al. 2006).  It is
possible that all ICMEs are magnetic clouds if viewed appropriately
(Krall 2007). This suggestion is consistent with the ubiquitous
nature of post eruption arcades, which seem to indicate flux rope
formation in the eruption process (Kang et al. 2006; Qiu et al.
2007; Yurchyshyn 2008).  While the reconnection process certainly
forms a flux rope, it is not clear if the reconnection creates a new
flux rope or fattens an existing one.

\subsection{SEP Events}

Energetic storm particle (ESP) events are the strongest evidence for
SEP acceleration by shocks, but this happens when the shocks arrive
at the observing spacecraft near Earth (Cohen et al. 2006). This
means the shocks must have been stronger near the Sun accelerating
particles to much higher energies. The strongest evidence for SEPs
in flares is the gamma-ray lines, which are now imaged by RHESSI
(Lin 2007). All shock-producing CMEs are associated with major
flares (M- or X-class in soft X-rays), so both mechanisms must
operate in most SEP events. There has been an ongoing debate as to
which process is dominant based on SEP properties such as the
spectral and compositional variability at high energies (Tylka \&
Lee 2006; Cane et al. 2007).

The easiest way to identify shocks near the Sun are the type II
radio bursts especially at frequencies below 14 MHz, which
correspond to the near-Sun IP medium (Gopalswamy 2006c). Analyzing
electrons and protons in SEP events, Cliver \& Ling (2007) have
found evidence for a dominant shock process including flatter SEP
spectra, apparent widespread sources, and high association with long
wavelength type II bursts. A recent statistical study finds the SEP
association rate of CME steadily increases with CME speed and width
especially and there is one-to-one correspondence between SEP events
and CMEs from the western hemisphere with long wavelength type II
bursts (Gopalswamy et al. 2008b).  Type II burst studies have also
have concluded that the variability in Alfven speed in the outer
corona decides the formation and strength of shocks (Shen et al.
2007; Gopalswamy et al. 2008c). For example, a 400 km s$^{-1}$ CME
can drive a shock, while a 1000 km s$^{-1}$ CME may not drive a
shock, depending on the local Alfven speed.

\vspace{3mm}

{\hfill James A. Klimchuk}

{\hfill {\it President of the Commission}}

\end{document}